\documentstyle[12pt,epsfig]{article}
\newcommand{\be}[1]{\begin{equation} \label{(#1)}}
\newcommand{\ee}{\end{equation}}
\newcommand{\ba}[1]{\begin{eqnarray} \label{(#1)}}
\newcommand{\ea}{\end{eqnarray}}
\newcommand{\nn}{\nonumber}

\def\znbb{0\nu\beta\beta}

\def\nrightarrow{\rightarrow\hspace{-1em}/\ \ }

\begin{document}
\hfill{USM-TH-193}\\[0.5cm]
%\hfill{IFIC-06/xx}
\begin{center}

{\Large\bf Extended Black Box Theorem for Lepton Number and Flavor Violating
processes}\\[3mm]
Martin Hirsch, $^1$ Sergey Kovalenko $^2$ and  Ivan Schmidt $^2$\\[1mm]
{\it $^1$ AHEP Group, Institut de F\'{\i}sica Corpuscular \\
  C.S.I.C./Universitat de Val{\`e}ncia \\
  Edificio Institutos de Paterna, Apt 22085, E--46071 Valencia, Spain \\[2mm]
$^2$ Departamento de F\'\i sica, Universidad
T\'ecnica Federico Santa Mar\'\i a,}\\
{\it Casilla 110-V, Valpara\'\i so, Chile}
\end{center}
\bigskip

\begin{abstract}
We revisit the well known ``Black Box" theorem establishing a
fundamental relation between the amplitude of neutrinoless double
beta decay and the effective Majorana neutrino mass. We extend this
theorem to the general case of arbitrary lepton number and lepton
flavor violating (LFNV) processes and to the three generation
Majorana neutrino mass matrix. We demonstrate the existence of a
general set of one-to-one correspondence relations between the
effective operators generating these processes, and elements of the
neutrino mass matrix, such that if one of these two quantities
vanishes the other is guaranteed to vanish as well, and moreover, if
one of these quantities is non-zero the other is guaranteed to be
non-zero. We stress that this statement remains valid even in the
presence of arbitrary new physics contributions. As a particularly
important example, we then show that neutrino oscillation data imply
that neutrinoless double beta decay must occur at a certain non-zero
rate.

\end{abstract}
\vskip 0.5cm

\bigskip
\bigskip


\bigskip
\bigskip


\newpage

\section{Introduction}

In the past few years neutrino oscillation experiments \cite{Fukuda:1998mi}
have, for the first time, provided unambiguous evidence for lepton flavor
violation (LFV). The current status of the experimental data can be briefly
summarized as follows \cite{Maltoni:2004ei}. Two neutrino mass squared
differences and two neutrino angles are known to be non-zero. These
are the atmospheric neutrino mass, $\Delta m^2_{\rm Atm} = (2.0-3.2)$
[$10^{-3}$ eV$^2$], and angle, $\sin^2\theta_{\rm Atm} = (0.34-0.68)$,
as well as the solar neutrino mass $\Delta m^2_{\odot} = (7.1-8.9)$
[$10^{-5}$ eV$^2$], and angle, $\sin^2\theta_{\odot} = (0.24-0.40)$,
all numbers at 3 $\sigma$ c.l. For the remaining neutrino angle,
often called the reactor neutrino angle $\theta_R$, and for the overall
neutrino mass scale only upper limits exist \cite{Maltoni:2004ei,PDG}.

Interestingly, the observed LFV can not exist isolated only in the neutrino
sector. It inevitably is carried over to the sector of charged leptons via
the LFV charged current loop with virtual neutrinos and should manifest
itself in the form of LFV processes with charged leptons, such as $\mu$,
$\tau$ and meson LFV decays, $\mu^--e^-$ nuclear conversion, etc. Thus,
given the observation of neutrino oscillations, the existence of yet
unobserved LFV processes with charged leptons is expected.

However, the LFV in the charged lepton sector may also receive
direct contributions from non-SM interactions of new physics. These
may cancel the contribution of the loop induced effect related to
neutrino oscillations. Such a cancellation is phenomenologically
unnatural, but possible if there exists some symmetry protecting
this cancellation. This observation gives rise to the question as to
whether the observation of neutrino oscillations guarantees non-zero
rates of some LFV processes with charged leptons and vice versa. In
view of the experimental observation of neutrino oscillations the
latter part of this question may be considered a purely theoretical
curiosity but it may reveal some generic relations between the
neutrino mass matrix and LFV processes with charged leptons.

One particular example of such kind of relation has been known for
some time. It is generally referred to as the ``Black Box'' theorem
\cite{BB}, and relates the effective Majorana neutrino mass and the
amplitude of neutrinoless double beta decay. In essence, it
establishes a one-to-one correspondence between these two
quantities, such that if one of them vanishes the other is
guaranteed to vanish and if one of these quantities is non-zero the
other must be non-zero as well. \footnote{A supersymmetric version
of the ``Black Box'' theorem has been formulated in Ref.
\cite{SUSY-BB}, extending this relation also to the lepton number
violating scalar neutrino mass.} The crucial point of the ``Black
Box'' theorem is that it remains valid in the presence of arbitrary
contributions of new physics. Note, however, that the existing
version of the ``Black box'' theorem \cite{BB} is limited to the
relation between neutrinoless double beta decay and the Majorana
property of the $\nu_e$ neutrino flavor, without properly taking
into account its mixing with (at least two) other neutrino flavor
states $\nu_{\mu}$ and $\nu_{\tau}$.

In the present paper we therefore revisit the ``Black Box'' theorem
and extend it to the general case of arbitrary lepton number and
flavor violating (LFNV) processes and the three-generation Majorana
neutrino mass matrix. Applying symmetry arguments we will show that
there exists a one-to-one correspondence between the amplitudes of
the LFNV processes and the corresponding entries of the Majorana
neutrino mass matrix. The same symmetry arguments allow us to
formulate self-consistency conditions for the Majorana neutrino mass
matrix, thus restricting its allowed index structure. We study in
more details the specific case of neutrinoless double beta decay and
the corresponding $M^{\nu}_{ee}$ element of the Majorana neutrino
mass matrix. We show that the self-consistency condition, in
combination with neutrino oscillation data, exclude the case
$M^{\nu}_{ee}=0$, guaranteeing that neutrinoless double beta decay
must occur at a non-zero rate.

\section{LFNV Processes and Neutrino mass Matrix}

Here we are going to prove a generic set of one-to-one correspondence
relations between LFV processes with total lepton number violation (LFNV)
$\Delta L = \Delta (L_{\alpha} + L_{\beta}) = 2$ and the corresponding
entries of the Majorana neutrino mass matrix. We consider baryon number
conserving processes in which LFNV manifests itself via two external
charged leptons. In addition, these processes may contain any set of
external SM particles satisfying certain conditions to be specified below.
For convenience we put all these additional particles to the initial states
of the LFNV processes while the two charged leptons appear in the final
states. Nevertheless our conclusions are valid for any $\Delta L = 2$
process, as will be seen later on.

Without loss of generality we focus on $\Delta L = 2$ processes of the
form (see Fig 1(a).): $\Phi_k\rightarrow l_{\alpha} l_{\beta}$, where
$\Phi_k$ denotes certain color-singlet subsets of external particles
with $B=0, L_{\alpha}=0$ and total electric charge $Q=-2$. The difference
between the processes with the same lepton flavor structure,
$l_{\alpha} l_{\beta}$, is due to the difference of their external
particle subsets $\Phi_k$ marked with different subscript $k$.
The possible subsets considered are of the following two types:
\begin{eqnarray}\label{set}
\Phi_0 &=& W^- W^-,\ \ \ \Phi_k =  (\bar{u} d)(\bar{u} d) \left\{(\bar{u}u),
(\bar{d}d), ... \right\}
\left\{(\bar{l}_{\alpha} l_{\alpha}), \gamma, Z \right\}.
\end{eqnarray}
The second type of sets must contain two pairs of anti-up and down quarks
of arbitrary generation in order to provide $Q=-2$. They may or may not
be accompanied by an arbitrary neutral set of particles.

We write down the effective Lagrangian describing the LFNV processes in 
the following schematic form:
\begin{eqnarray}\label{eff-Lag-1}
{\cal L}^{\Delta L=2} &=&
             \Phi_k \bar{l}_{\alpha} \Gamma^{(k)}_{\alpha\beta} l^c_{\beta}  +
        g \bar{l}_{\alpha}\gamma^{\mu}P_L \nu_{\alpha} W_{\mu}^-
             + \Phi_k W^-_{\mu}S^{\mu\nu}_k W^-_{\nu}
            + {\cal L}^{\prime}  + h.c.
\end{eqnarray}
Here, $P_L = (1-\gamma_5)/2$.  The first effective operator
generates amplitudes for $\Phi_k\rightarrow l_{\alpha} l_{\beta}$
processes and may receive contributions from SM interactions with
Majorana neutrino mass insertions (to be discussed in the next
section) and possibly from some physics  beyond the SM. The vertex
structure $\Gamma^{(k)}$ depends on the concrete realization of this
operator. The second term is the usual SM leptonic charged current
interaction. The third term represents the SM interactions of the
sets of fields $\Phi_k$ with W-bosons.  This term plays an important
role in our analysis and is always induced by the SM interactions.
This is guaranteed by the structure of the field set $\Phi_k$ given
in Eq. (\ref{set}) and directly follows from the presence of the two
$\bar{u}d$ pairs which have SM couplings to $W^-$. The term ${\cal
L}^{\prime}$ denotes any interactions beyond the SM, whose explicit
form is irrelevant for our analysis. The particular realization of
the above terms depends on the field set $\Phi_k$ and determines the
vertex structure $S_k$. The explicit form of $\Gamma^{(k)}$ and
$S_k$ are also irrelevant for our subsequent reasoning. In what
follows we denote the first and second effective operators in Eq.
(\ref{eff-Lag-1}) by $\hat{\Gamma}^{(k)}_{\alpha\beta}$ and
$\hat{S}_k$, respectively.

Majorana mass terms for the three left handed neutrinos
$\nu_{\alpha}=\nu_{e},\nu_{\mu},\nu_{\tau}$ in 4-component notations
can be written as
\begin{eqnarray}\label{nu-mass-matrix}
{\cal L}^{M}_{\alpha\beta} = -\frac{1}{2}
\overline{\nu_{\alpha}^c} M^{\nu}_{\alpha\beta} P_L \nu_{\beta} + \mbox{h.c.}
\end{eqnarray}

It is obvious that neutrinos contribute to the operator
$\hat{\Gamma}^{(k)}_{\alpha\beta}$ in Eq. (\ref{eff-Lag-1}) via the
well known diagram in Fig. 1(a), leading to the following expression
for the vertex function
\begin{eqnarray}\label{nu-contribution}
\Gamma^{(k)}_{\alpha\beta}(x_1-x_2,...) &\sim&
\int \frac{d^4q}{(2\pi)^4} M^{\nu}_{\alpha m}
\left(q^2 - M^{\nu\dagger}M^{\nu}\right)_{m\beta}^{-1}
e^{-iq(x_1 - x_2)} = \\ \nn
&=& \int \frac{d^4q}{(2\pi)^4} q^{-2}M^{\nu}_{\alpha m} \left(\delta_{m\beta} +
\sum_{n=1}^{\infty} q^{-2n}
\left(M^{\nu\dagger}M^{\nu}\right)_{m \beta}^{n}\right) e^{-iq(x_1 - x_2)}.
\end{eqnarray}
Here we have formally expanded the denominator in the ratio
$M_{(\nu)}^2/q^2$. The terms of this expansion correspond, in the
flavor basis, to the contributions of all possible mass insertions in the 
intermediate neutrino line, see Fig.1(a). The number of insertions must be
odd in order to have total lepton number violation $\Delta L =2$,
since each Majorana mass insertion introduces $\Delta L = \pm 2$.
Thus, the leading order neutrino contribution to the operator
$\hat{\Gamma}^{(k)}_{\alpha\beta}$  is proportional to the
$M^{\nu}_{\alpha\beta}$ entry of the Majorana neutrino mass matrix.
However, this operator receives higher order contributions from the
other entries which can generate it even in the absence of an
$M^{\nu}_{\alpha\beta}$ entry. The infinite serie in Eq. 
(\ref{nu-contribution}) can be represented in a compact form on the basis 
of the Hamilton-Kelly theorem \cite{Hamilton-Kelly},
which asserts that any analytic function of an
$n\times n$ matrix is equivalent to an n-1 order polynomial of this matrix.   
Applying this theorem to Eq. (\ref{nu-contribution}) we get
\begin{eqnarray}\label{H-K}
\Gamma^{(k)}_{\alpha\beta} &=& c_0 M^{\nu}_{\alpha\beta} +  c_1
M^{\nu}_{\alpha k}\left(M^{\nu\dagger}M^{\nu}\right)_{k\beta} +
 c_2 M^{\nu}_{\alpha k}\left(M^{\nu\dagger}M^{\nu}\right)^2_{k\beta}
+ \Gamma^{\prime}\, .
\end{eqnarray}
The coefficients $c_n$ depend on the mass eigenvalues $m^2_{\nu_{1,2,3}}$ 
of the matrix $M^{\nu\dagger}M^{\nu}$. Here, in addition to the first 
three terms following from Eq. (\ref{nu-contribution}) we allowed for a 
term $\Gamma^{\prime}$ representing any possible contributions other 
than those of Fig. 1(a). The above formula represents the net result of 
all possible mass insertions. 
The advantage of the representation (\ref{H-K}) is that it specifies all
the irreducible sets of non-trivial LFV transitions, into which any other
LFV transition in the infinite sum of Eq. (\ref{nu-contribution}) can be
decomposed.  We will use this fact later.

Now consider the contribution of the operator
$\hat{\Gamma}^{(k)}_{\alpha\beta}$ in Eq. (\ref{eff-Lag-1}) to the Majorana
neutrino mass matrix $M^{\nu}_{\alpha\beta}$ in Eq. (\ref{nu-mass-matrix}).
This contribution is generated by the 1PI self-energy diagrams in Fig. 1(b)
with two SM charged current vertices. In leading order, taking into account
all non-trivial lepton flavor transitions, we write the corresponding
contribution to the Majorana neutrino mass matrix in a form similar to
that of Eq. (\ref{H-K}):
\begin{eqnarray}\label{O-M}
M^{\nu}_{\alpha\beta} &=& \Sigma_0\langle \Gamma_{\alpha\beta}\rangle +
\Sigma_1\langle \Gamma_{\alpha m} (\Gamma^{\dagger}\Gamma)_{m\beta} \rangle
+ \Sigma_2\langle \Gamma_{\alpha m} (\Gamma^{\dagger}\Gamma)^2_{m\beta}\rangle
+ M^{\prime},
\end{eqnarray}
where in the brackets $\langle \rangle$ a convolution in coordinate space
and matrix multiplication in flavor space are implied. As in Eq. (\ref{H-K}),
here we also allowed for a term $M^{\prime}$ representing any possible
contributions other than in Fig. 1(b). The explicit form of the terms in
Eqs. (\ref{O-M}),(\ref{H-K}) is not important for our subsequent analysis.
Decisive, however, is the observation that the operator
$\hat{\Gamma}^{(k)}_{\alpha\beta}$ from Eq. (\ref{eff-Lag-1}) and the
Majorana neutrino mass matrix $M^{\nu}_{\alpha\beta}$ defined in
Eq. (\ref{nu-mass-matrix}) mutually contribute to each other.

\section{Extended ``Black Box Theorem"}

Now we are in a position to prove the Extended ``Black Box'' Theorem
establishing a one-to-one correspondence between the amplitudes of
LNFV processes and the entries of the Majorana neutrino mass matrix.
First, let us explore the consequences of the assumption
$M^{\nu}_{\alpha\beta}=0$ for $\hat{\Gamma}^{(k)}_{\alpha\beta}$ and
vice versa $\hat{\Gamma}^{(k)}_{\alpha\beta}=0$ for
$M^{\nu}_{\alpha\beta}$. To this end let us consider Eq. (\ref{O-M})
for the former case and Eq. (\ref{H-K}) for the latter one. The
condition $M^{\nu}_{\alpha\beta}=0$ in Eq. (\ref{O-M}) implies that
the sum of the terms on the r.h.s. must vanish. This can happen
either if each of the four terms vanish individually or if these
four non-zero terms cancel each other. The latter case of an
accidental self-cancellation is unnatural, since it is in general
unstable with respect to radiative corrections. The same arguments
can be applied to Eq. (\ref{H-K}). Thus, on the basis of naturalness
arguments alone one may expect from Eqs. (\ref{H-K}) and (\ref{O-M})
that $M^{\nu}_{\alpha\beta}=0$ requires
$\hat{\Gamma}^{(k)}_{\alpha\beta}=0$ and vice versa. This conclusion
may be circumvented only if there exists a symmetry protecting the
cancellation to all orders of perturbation theory.

Let us examine if any global symmetry of this type in a theory with the
Lagrangian Eq. (\ref{eff-Lag-1}), which includes both SM and beyond the SM
interactions, can exist. Such a symmetry should be associated with a
unitary transformation of the fields realizing representations of a global
group, which we denote by $G_{\eta}$. Suppose under this symmetry the fields
transform as
\begin{eqnarray}\label{group}
W^- \stackrel{G_\eta}{\longrightarrow} \eta_{_W} \cdot W^-,\ \
l_{\alpha} \stackrel{G_\eta}{\longrightarrow} \eta^{l}_{\alpha}
\cdot l_{\alpha}, \ \ \
\nu_{\alpha} \stackrel{G_\eta}{\longrightarrow} \eta^{\nu}_{\alpha}
\cdot \nu_{\alpha}, \ \ \
\Phi_k \stackrel{G_\eta}{\longrightarrow} \eta_k \cdot \Phi_k,
\end{eqnarray}
with $\eta_a \eta_a^{\dagger} = 1$. Terms in Eq. (\ref{eff-Lag-1})
will then be allowed (forbidden) by the symmetry if the corresponding
product of $\eta$ factors is equal to $1$ (different from $1$).
The group $G_{\eta}$ is not
completely arbitrary since it must at least be consistent with the
SM part of the Lagrangian (\ref{eff-Lag-1}). The invariance of the second
and third terms in Eq. (\ref{eff-Lag-1}) with respect to $G_{\eta}$
requires:
\begin{eqnarray}\label{constr}
\eta_{_W}^2\eta_k =1, \ \ \ \  \eta_{_W} \eta_{\alpha}^{l \dagger}
\eta^{\nu}_{\alpha} =1.
\end{eqnarray}
The effective operators $\hat{\Gamma}^{(k)}_{\alpha\beta}$ generating the
processes $\Phi_k\rightarrow l_{\alpha} l_{\beta}$ and the Majorana neutrino
mass terms transform under $G_{\eta}$ as
\begin{eqnarray}\label{Gamma1}
\hat{\Gamma}^{(k)}_{\alpha\beta} &\stackrel{G_\eta}{\longrightarrow}&
\eta^k_{\alpha\beta} \cdot \hat{\Gamma}^{(k)}_{\alpha\beta}, \ \ \
\mbox{with}\ \
\eta^k_{\alpha\beta} = \eta^{l}_{\alpha}\cdot\eta^{l}_{\beta}\cdot \eta_k,\\
\label{Gamma1-1}
{\cal L}^{M}_{\alpha\beta} &\stackrel{G_\eta}{\longrightarrow}&
\eta^{\nu}_{\alpha\beta} \cdot {\cal L}^{M}_{\alpha\beta}, \ \ \ \ \
\mbox{with}\ \
\eta^{\nu}_{\alpha\beta} = \eta^{\nu}_{\alpha}\cdot\eta^{\nu}_{\beta}.
\end{eqnarray}
From condition (\ref{constr}) it follows that
$\eta_k = (\eta_{_W}^{\dagger})^2$ and
$\eta^{l}_{\alpha} = \eta^{\nu}_{\alpha}\eta_{_W}$ and thus
\begin{eqnarray}\label{Gamma2}
\eta^k_{\alpha\beta} = \eta^{\nu}_{\alpha\beta}\ \ \
\mbox{for}\ \ \ \forall\  k\ .
\end{eqnarray}
We repeat that terms in the Lagrangian Eqs. (\ref{eff-Lag-1}) and
(\ref{nu-mass-matrix}) are allowed by $G_{\eta}$ if the corresponding
transformation factor of this term is $\eta=1$ and forbidden if $\eta\neq1$.
\footnote{Similar arguments have been used previously to prove a one-to-one
correspondence between the amplitude of neutrinoless double beta decay,
Majorana neutrino \cite{BB} and sneutrino \cite{SUSY-BB} masses as well as
between different LFNV processes \cite{LFV-rel}.}
Eq. (\ref{Gamma2}) proves a general one-to-one correspondence between
the effective operators $\hat{\Gamma}^{(k)}_{\alpha\beta}$, generating
the processes $\Phi_k\rightarrow l_{\alpha} l_{\beta}$, and the elements
of the neutrino mass matrix $M^{\nu}_{\alpha\beta}$.
According to Eq. (\ref{Gamma2}), the effective operator
$\hat{\Gamma}^{(k)}_{\alpha\beta}$ and the corresponding element of the
mass matrix $M^{\nu}_{\alpha\beta}$ are related such that if one of these
two quantities vanishes the other is guaranteed to vanish, and vice versa
if one of these quantities is non-zero the other one is non-zero as well.
The crucial point of this statement is that it is valid in the presence
of any contributions of new physics represented in Eqs. (\ref{eff-Lag-1}),
(\ref{H-K}) and (\ref{O-M}) by ${\cal L}^{\prime}$, $\Gamma^{\prime}$
and $M^{\prime}$. Schematically one can express these relations as:
\begin{eqnarray} \label{rel}
(\Phi_k\nrightarrow l_{\alpha} l_{\beta}) \ \hat{\Gamma}^{(k)}_{\alpha\beta}
= 0\   \leftrightarrow \  M^{\nu}_{\alpha\beta}=0, \ \ \
(\Phi_k\rightarrow l_{\alpha} l_{\beta}) \ \hat{\Gamma}^{(k)}_{\alpha\beta}
\neq 0\    \leftrightarrow \  M^{\nu}_{\alpha\beta} \neq 0, \ \ \
\end{eqnarray}
This result is valid not only for the processes of type
$\Phi_{k}\rightarrow l_{\alpha} l_{\beta}$ but for any
$\Delta L = \Delta (L_{\alpha} + L_{\beta}) = 2$ process like
$\Phi^{(1)}_{k}\rightarrow \Phi^{(2)}_{n} l_{\alpha} l_{\beta}$,
$\Phi^{(1)}_{k} \bar{l}_{\alpha} \rightarrow \Phi^{(2)}_{n}  l_{\beta}$,
because our conclusions are based on the analysis of the effective
operator $\hat{\Gamma}^{(k)}_{\alpha\beta}$, which generates all these
processes.

Eq. (\ref{rel}) represents a direct extension of the flavor blind
``Black Box'' Theorem \cite{BB} to the case involving arbitrary
LNFV. However, due to the possibility of multiple LNFV transitions,
in the latter case there are additional indirect relations between
$\Phi_k\rightarrow l_{\alpha} l_{\beta}$ and $M^{\nu}_{kn}$ for
$k\neq \alpha, n\neq \beta$. Let us turn again to Eqs. (\ref{H-K})
and (\ref{O-M}). The relations Eq. (\ref{rel}) correspond to the
leading order terms on the r.h.s. of Eqs. (\ref{H-K}) and
(\ref{O-M}) when $\hat{\Gamma}^{(k)}_{\alpha\beta}$ and
$M^{\nu}_{\alpha\beta}$ are directly proportional to each other.
However, according to Eq. (\ref{H-K})  and (\ref{O-M}), the second
and third (next-to-leading) terms introduce new relations between
$\hat{\Gamma}^{(k)}_{\alpha\beta}$ and $M^{\nu}_{mn}$ for $m\neq
\alpha, n\neq \beta$. One can find all these residual relations
using the above symmetry arguments. First, using Eqs.
(\ref{Gamma1})-(\ref{Gamma2}), the following relations for the
$G_{\eta}$ symmetry factors are obtained
\begin{eqnarray}\label{next}
&&\eta^{\nu}_{\alpha\beta} = \eta^k_{\alpha n}\eta^{k\ \dagger}_{mn}
\eta^k_{m\beta}, \ \ \ \ \ \ \ \ \ \ \
\eta^{k}_{\alpha\beta} = \eta^{\nu}_{\alpha n}\eta^{\nu\ \dagger}_{mn}
\eta^{\nu}_{m\beta},\\
\label{next1}
&&\eta^{\nu}_{\alpha\beta} = \eta^k_{\alpha n}\eta^{k\ \dagger}_{mn}
\eta^k_{ml}\eta^{k\ \dagger}_{pl}\eta^k_{p\beta}, \ \ \
\eta^{k}_{\alpha\beta} = \eta^{\nu}_{\alpha n}\eta^{\nu\ \dagger}_{mn}
\eta^{\nu}_{ml}\eta^{\nu\ \dagger}_{pl}\eta^{\nu}_{p\beta},
\end{eqnarray}
These relations follow from the fact that $\eta^k_{mn}=\eta^\nu_{mn}
=\eta^\nu_{m}\eta^\nu_{n}$ and $\eta^\nu_{i}\eta^{\nu \dagger}_{i}=1$.  
In principle, one can write down an infinite set of such relations involving
multiple LNFV transitions corresponding to the mass insertions in
Fig. 1(a), given by the infinite series in Eq. (\ref{nu-contribution}) as well
as the self-energy insertions to the massless neutrino
propagator of Fig. 1(b). However, as we already have noted the irreducible 
set of transitions is given by the first three terms of Eqs.
(\ref{H-K}) and (\ref{O-M}), as follows from the Hamilton-Kelly
theorem. Eqs. (\ref{Gamma2}), (\ref{next}) and (\ref{next1}) represent a 
complete set of relations between the transformation  $\eta$-factors 
corresponding to these terms. In fact, let us consider Eq. (\ref{H-K}) 
and observe that the field transformations in Eqs. (\ref{group}) are 
equivalent to the transformations of the vertex function and 
neutrino mass matrix:
%
%\begin{eqnarray}\label{equiv}  
$\Gamma^{(k)}_{\alpha\beta} \stackrel{G_\eta}{\longrightarrow}
\eta^k_{\alpha\beta} \cdot \Gamma^{(k)}_{\alpha\beta}, \ 
M^{\nu}_{\alpha\beta} \stackrel{G_\eta}{\longrightarrow}
\eta^{\nu}_{\alpha\beta} \cdot M^{\nu}_{\alpha\beta}$. 
%\end{eqnarray}
Both sides of Eq. (\ref{H-K}) must equally transform under $G_{\eta}$. 
The necessary and sufficient conditions to satisfy this requirement 
are given by Eqs. (\ref{Gamma2}), (\ref{next}) and (\ref{next1}). In 
other words they represent a complete set of non-trivial relations 
between the symmetry $\eta$-factors. Another way to demonstrate the 
completeness of the relations Eqs. (\ref{Gamma2}), (\ref{next}) and 
(\ref{next1}) is given by combinatorial analysis. In the three generation 
case under consideration it is easy to show that any relationships other 
than the type given in Eqs. (\ref{next})-(\ref{next1}) can be decomposed 
into blocks corresponding to irreducible sets of $\eta$-factors on the
r.h.s. of these equations. 

Thus it is sufficient to consider Eq. (\ref{next})-(\ref{next1}), together 
with Eq. (\ref{Gamma2}), in order to find all possible relations
between $\hat{\Gamma}^{(k)}$  and $M^{\nu}$, following from our symmetry 
arguments. The information provided by Eqs. (\ref{next})-(\ref{next1}) 
can be expressed in various ways. We
formulate the consequences of these equations in the form of the
following question: what are the minimal sets of the LFNV processes
to be observed experimentally that are sufficient to prove that
$M^{\nu}_{\alpha\beta}\neq 0$ ? Naturally, the observation of
$\Phi_{k}\rightarrow l_{\alpha} l_{\beta}$ would establish this fact
as evidenced by Eqs. (\ref{rel}). However, there exist other sets of
processes $\Phi_{k}\rightarrow l_m l_n$ with $m\neq \alpha$ and/or $n\neq
\beta$, whose observation would also establish
$M^{\nu}_{\alpha\beta}\neq 0$. The complete lists of such processes
corresponding to each entry of $M^{\nu}_{\alpha\beta}$ can be
derived from Eqs. (\ref{next})-(\ref{next1}). Here for illustration
we only show the complete list of LFNV processes necessary to
establish that $M^{\nu}_{ee}\neq 0$. In this case there are five
independent sets of experiments shown in the curl brackets:
\begin{eqnarray}\label{list}
M^{\nu}_{ee}\neq 0 \longleftarrow
\left\{
\begin{array}{l}
\left\{(e\mu),(\mu\mu)\right\},\ \left\{(e\tau),(\tau\tau)\right\},
\ \left\{(e\mu),(\mu\tau),(e\tau)\right\}, \\
\left\{(e\mu),(\mu\tau),(\tau\tau)\right\},\ \left\{(e\tau),(\mu\tau),
(\mu\mu)\right\}.
\end{array}
\right.
\end{eqnarray}
Here we denoted the process $\Phi_{k}\rightarrow l_{m} l_{n}$
by $(mn)$, where $m\neq e$ and/or $n\neq
e$ . The first three sets of experiments follow from Eq. (\ref{next})
and the last two from Eq. (\ref{next1}). For instance, the observation of 
both $\Phi_{k}\rightarrow l_{e} l_{\mu}$ and 
$\Phi_{k}\rightarrow l_{\mu} l_{\mu}$ would establish $M^{\nu}_{ee}\neq 0$, 
etc. In practice, information of this type might become
useful in a situation when for some (experimental) reasons the observation
of certain $\Phi_{k}\rightarrow l_{\alpha} l_{\beta}$ processes is not 
possible. In the above example this is $\Phi_{k}\rightarrow l_{e} l_{e}$.

Formally the Extended Black Box Theorem is formulated as Eqs. (\ref{Gamma2})
and (\ref{next})-(\ref{next1}), allowing one to extract various relations 
of the above discussed type between LFNV processes and entries of the 
Majorana neutrino mass matrix $M^{\nu}$.

%%%%%%%%%%%%%%%%%%%%%%%%%%%%%%%%%%%%%%%%%%%%%%%%%%

\section{Neutrinoless double beta decay}

We will now apply the symmetry arguments of the previous section
to analyze the structure of the Majorana neutrino mass matrix
itself. As the phenomenologically most relevant lepton number
violating process, we will concentrate on the consequences for
neutrinoless double beta decay ($\znbb$). As is well-known, Majorana
neutrino masses induce a $\znbb$ decay amplitude proportional to
\begin{eqnarray}\label{meff}
\langle m_{\nu}\rangle & = & \sum_j U_{ej}^2 m_j \\
& \equiv & M^{\nu}_{ee}
\end{eqnarray}
Here, the sum includes all light neutrino mass eigenstates. The
second relation, i.e. that $\znbb$ decay is sensitive to the
($\nu_e-\nu_e$) element of the neutrino mass matrix in flavor space,
follows from a straightforward calculation.

From the discussion presented above it is clear that $M^{\nu}_{ee} \equiv 0$
requires $\eta^{\nu}_{ee} \ne 1$. We will now show that $\eta^{\nu}_{ee}
\ne 1$ is inconsistent with data from neutrino oscillation experiments.
The argument is essentially a proof by contradiction: construct all
possible neutrino mass matrices consistent with $\eta^{\nu}_{ee} \ne 1$;
if none of the resulting matrices can explain all oscillation data,
$M^{\nu}_{ee}$ must be non-zero.

As an example, assume $\eta^{\nu}_{ee} = \eta^{\nu}_{e}\eta^{\nu}_{e}\ne 1$.
Choose $\eta^{\nu}_{e\mu}=\eta^{\nu}_{e}\eta^{\nu}_{\mu} = 1$ and
$\eta^{\nu}_{e\tau}=\eta^{\nu}_{e}\eta^{\nu}_{\tau} = 1$. It follows
that $\eta^{\nu}_{\mu}= (\eta^{\nu}_{e})^*$ and
$\eta^{\nu}_{\tau}= (\eta^{\nu}_{e})^*$, as well as
$\eta^{\nu}_{\mu\mu}\ne 1$, $\eta^{\nu}_{\mu\tau}\ne 1$ and
$\eta^{\nu}_{\tau\tau}\ne 1$. Thus the resulting mass matrix has the
structure \footnote{We stress that the zeros in these matrices are exact,
since we assume they are enforced by a symmetry.}
\begin{equation}\nonumber
{\cal M}_{\nu}^{(1)} =
\left(\begin{array}{cccc}
0 & x & y \\
x & 0 & 0 \\
y & 0 & 0
\end{array}\right) .
\end{equation}
In a completely analogous way one can find all possible neutrino mass 
matrices consistent with $M^{\nu}_{ee}=0$ and with a maximal number of 
non-zero entries. In addition to ${\cal M}_{\nu}^{(1)}$, the other 
possibilities have the following structures
\begin{equation}\nonumber
{\cal M}_{\nu}^{(2)} =
\left(\begin{array}{cccc}
0 & 0 & 0 \\
0 & x & y \\
0 & y & z
\end{array}\right)  \hskip5mm
{\cal M}_{\nu}^{(3)} =
\left(\begin{array}{cccc}
0 & x & 0 \\
x & 0 & y \\
0 & y & 0
\end{array}\right)
\end{equation}

\begin{equation}\nonumber
{\cal M}_{\nu}^{(4)} =
\left(\begin{array}{cccc}
0 & 0 & y \\
0 & x & 0 \\
y & 0 & 0
\end{array}\right) \hskip5mm
{\cal M}_{\nu}^{(5)} =
\left(\begin{array}{cccc}
0 & x & 0 \\
x & 0 & 0 \\
0 & 0 & y
\end{array}\right)
\end{equation}
As promised, none of these five matrices is consistent with the
data. ${\cal M}_{\nu}^{(1)}$ and ${\cal M}_{\nu}^{(3)}$ have
eigenvalues such that one $\Delta m^2 \equiv 0$. ${\cal M}_{\nu}^{(2)}$
leads to a solar angle $\theta_{\odot} \equiv 0$ and ${\cal M}_{\nu}^{(5)}$
predicts the atmospheric angle $\theta_{\rm Atm} \equiv 0$. And,
finally, ${\cal M}_{\nu}^{(4)}$ predicts the reactor angle to be
maximal and $\theta_{\odot} \equiv 0 \equiv \theta_{\rm Atm}$.

As we have already shown, $\znbb$ decay must occur at certain
non-zero rate if $M^{\nu}_{ee}\neq 0$ and is forbidden if
$M^{\nu}_{ee}\equiv 0$. We can conclude that, with the
assumption that neutrinos indeed are Majorana particles, current
neutrino oscillation data show that neutrinoless double beta decay
must occur with a non-zero rate.

\section{Conclusions}

We have proven a generic set of one-to-one correspondence relations
between LFV processes with total lepton number violation (LFNV)
$\Delta L = \Delta (L_{\alpha} + L_{\beta}) = 2$ and the
corresponding entries of the Majorana neutrino mass matrix. These
relations are an extension of the well known ``Black Box" theorem,
establishing a fundamental relation between the amplitude of
neutrinoless double beta decay and the effective Majorana neutrino
mass, to the general case of arbitrary lepton number and lepton
flavor violating (LFNV) processes and the three generation Majorana
neutrino mass matrix. As a particularly interesting application of
this theorem we have shown that neutrino oscillation data imply that
neutrinoless double beta decay must occur at a certain non-zero
rate.

%%%%%%%%%%%%%%%%%%%%%%%%%%%%%%%%%%%%%%%%%%%%%%%%%%%%%%%
\vskip15mm
\centerline{\bf Acknowledgments}

This work was supported by the FONDECYT projects 1030244 and
1030355, by Spanish grant FPA2005-01269 and by the European
Commission Human Potential Program RTN network MRTN-CT-2004-503369,
as well as the EU Network of Astroparticle Physics
(ENTApP) WP1. M.H. is supported by an MCyT Ramon y Cajal contract.

\bigskip
\newpage

\begin{figure}
\begin{center}
\epsfig{file=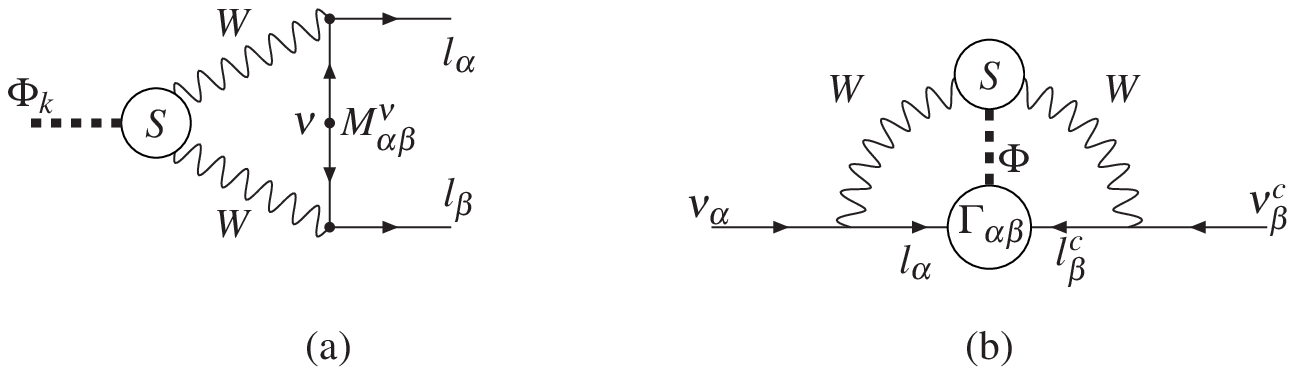}\\
\caption{(a) Contribution of the $M^{\nu}_{\alpha\beta}$ entry of the 
Majorana neutrino mass matrix to the effective LFNV vertex 
$\Gamma_{\alpha\beta}$ and  visa versa (b) contribution of 
$\Gamma_{\alpha\beta} $ to  $M^{\nu}_{\alpha\beta}$. }
\end{center}
\end{figure}

\end{document}